\documentclass[prl,twocolumn, aps,amssymb,footinbib,showpacs]{revtex4-1}

\usepackage{graphicx}
\usepackage{amsmath,amssymb}
\usepackage{cancel}
\usepackage{color}
\usepackage{pifont}
\usepackage{float}

\newcommand{\ket}[1]{\ensuremath{\left| #1 \right\rangle}}

\begin{document}

\title{Many-body Localization in Dipolar Systems}

\author{N. Y. Yao$^{1}$, C. R. Laumann$^{1,2,3}$, S. Gopalakrishnan$^{1}$, M. Knap$^{1,4}$, M. M\"{u}ller$^{5}$, E. A. Demler$^{1}$, M. D. Lukin$^{1}$}

\affiliation{$^{1}$Department of Physics, Harvard University, Cambridge, MA 02138, U.S.A.}
\affiliation{$^{2}$Perimeter Institute for Theoretical Physics, Waterloo, ON N2L 2Y5, Canada}
\affiliation{$^{3}$Department of Physics, University of Washington, Seattle, WA 98195, U.S.A.}
\affiliation{$^{4}$ITAMP, Harvard-Smithsonian Center for Astrophysics, Cambridge, MA 02138, USA}
\affiliation{$^{5}$The Abdus Salam International Center for Theoretical Physics, Strada Costiera 11, 34151 Trieste, Italy}

\begin{abstract}
Systems of strongly interacting dipoles offer an attractive platform to study many-body localized phases, owing to their long coherence times and strong interactions. 
We explore conditions under which such localized phases persist in the presence of power-law interactions and supplement our analytic treatment with numerical evidence of localized states in one dimension.
We propose and analyze several experimental systems that can be used to observe and probe such states, including ultracold polar molecules and solid-state magnetic spin impurities. 
\end{abstract}


\maketitle


Statistical mechanics is the framework that connects thermodynamics to the microscopic world. It hinges on the assumption of equilibration; when equilibration fails, so does much of our understanding.
In isolated quantum systems, this breakdown is captured by the phenomenon known as many-body localization (MBL) \cite{Anderson58,Fleishman80,Altshuler97,Basko06,Gornyi05,Burin06,Oganesyan07,Pal10,Znidaric08,Monthus10,Bardarson12,Vosk12,Iyer13,Serbyn13,Huse13,Serbyn13b,Huse13b,Pekker13,Vosk13,Bahri13,Chandran13,Bauer13,Swingle13,Schiulaz13}. 
Many-body localized phases conduct neither matter, charge nor heat. Moreover, they can exhibit symmetry breaking and topological order in dimensions normally forbidden by Mermin-Wagner-type arguments \cite{Huse13b,Chandran13}.
To date, none of these phenomena has been observed in experiments, in part because of the isolation required to avoid thermalization.

In this Letter, we investigate dilute dipolar systems as a platform for realizing MBL phases and studying the associated localization phase transition.
Our work is motivated by recent experimental advances that make it possible to produce and probe isolated, strongly interacting ensembles of disordered particles, as found  in systems ranging from trapped ions \cite{Korenblit12} and Rydberg atoms \cite{Ryabtsev10,Nipper12} to ultracold polar molecules \cite{Miranda11, ni08} and spin defects in solid state systems \cite{Neumann10b, Childress06, Dutt07, Neumann08}.
The presence of strong interactions in these systems underlies their potential for exploring physics beyond that of single particle Anderson localization \cite{Anderson58}. 
However, the power law decay of those interactions immediately raises the question: can localization persist in the presence of such long-range interactions?
Indeed, Anderson observed in his seminal paper that long-ranged \emph{hopping} $t \sim 1/r^\alpha$ delocalizes any putatively localized single-particle states for $\alpha \le d$, with $d$, the dimension of space.  
In what  follows, we consider the generalization of Anderson's criterion to the \emph{interacting} power-law regime and produce a necessary condition for localization with such interactions \cite{Burin06}. 
To support these considerations, we carry out extensive numerical analysis of power law interacting systems in $d=1$ spatial dimension. 
With this criterion in hand, we analyze the feasibility of observing MBL states in two complementary ultracold polar molecule proposals, wherein the power laws, interaction scales and dimensionality may be tuned. Finally, we generalize our analysis to solid-state systems where localization can be studied in the quantum dynamics of magnetic spin impurities. 


\begin{figure}
	\centering
		\includegraphics[width=\columnwidth]{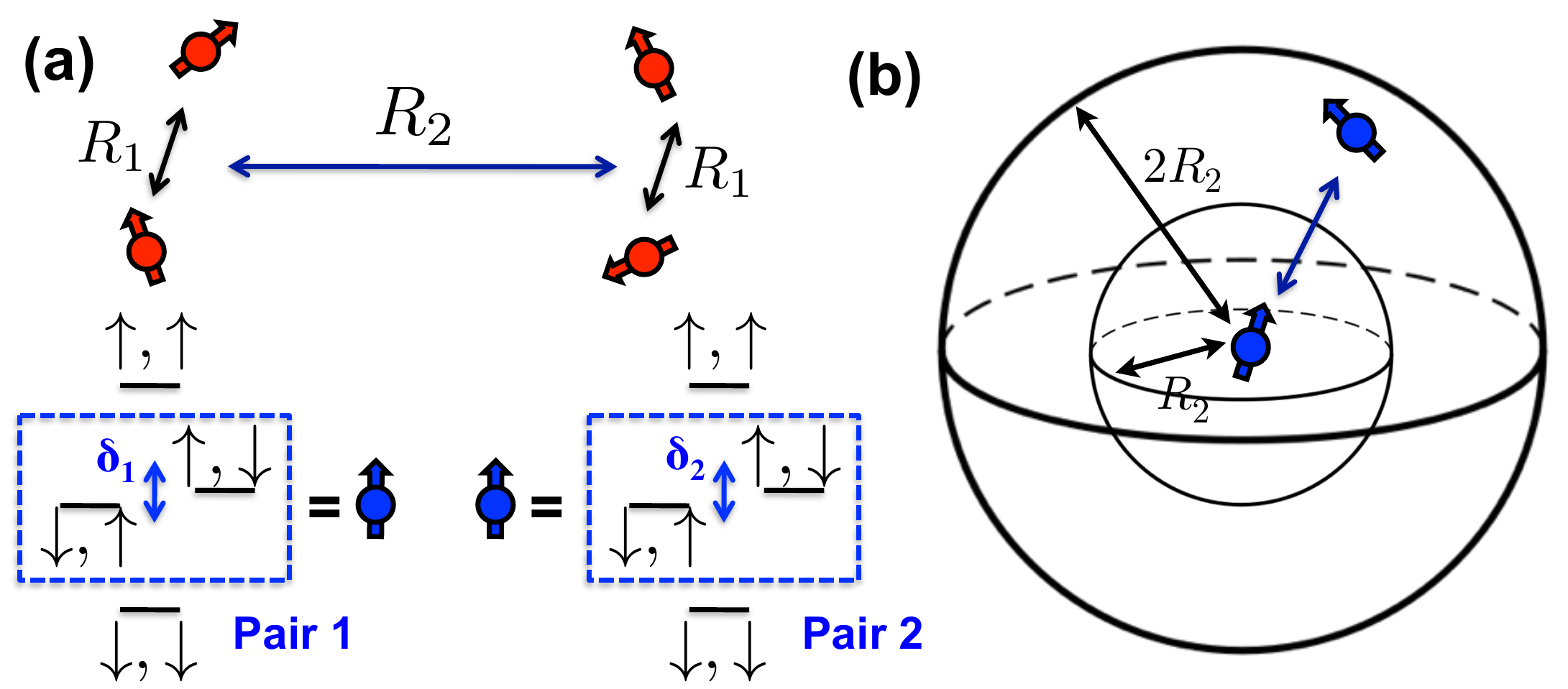}
	\caption{%
	(a) Schematic of four spin resonance structure. Each pair of (red) spins at separation $R$ forms a pseudospin (blue) with level structure shown below. 
	(b) A pseudospin at the origin resonates with another pseudospin in a shell $R'<r<2R'$.}
	\label{fig:fig1_v1_schem}
\end{figure}

\paragraph{Conditions for localization---}
In localized systems, injections of energy propagate at most a finite distance even after infinite time. 
This is obviously inconsistent with the proliferation of long-range resonances through which energy may be transported.
In the following, we identify resonant degrees of freedom and ask whether the number of such  resonances diverges at large scales;  such divergence  suggests the existence of a percolating network which conducts energy \cite{Burin06}. 
We consider a general two-body Hamiltonian of spin $1/2$ particles with conserved total $S^z$,
\begin{align}
	\label{eq:hamgeneral}
	H = \sum_i \epsilon_i S^z_i - \sum_{ij} \frac{t_{ij}}{|r_{ij}|^\alpha} (S^+_i S^-_j + h.c.) + \sum_{ij} \frac{V_{ij}}{|r_{ij}|^\beta} S^z_i S^z_j
\end{align}
where $\epsilon_i$ is a site dependent disorder field of bandwidth $W$, while $\alpha$ and $\beta$ are the exponents governing the power law decay of spin flip-flops and spin interactions, respectively  \cite{footnotetij}; we assume $\beta \le \alpha$ consistent with all  physical realizations of which we are aware. 
Clearly, the analysis applies to general long-range interacting two-level systems with a conserved charge.

We identify resonant pairs of spins as those for which $|\epsilon_i - \epsilon_j| \lesssim t_{ij}/|r_{ij}|^\alpha$; the expected number  of resonant spins at a distance $R_1 < |r_{ij}| < 2R_1$ from a central spin is
\begin{align}
	N_1(R_1) \sim (\rho R_1^d) \cdot \frac{t/R_1^\alpha}{W} 
	\label{eq:pairdensity}
\end{align} 
where $\rho$ is the density of spins.  
If $N_1(R_1)$ diverges as $R_1 \rightarrow \infty$, that is, if $d>\alpha$, then any spin resonates with arbitrarily distant spins and localization is impossible; this is precisely Anderson's criterion for single-particle localization. 
In the critical case, $d=\alpha$, a  detailed renormalization group treatment  confirms subdiffusive but delocalized behavior for the non-interacting case \cite{Anderson58,Levitov90,Levitov99}.

  \begin{table}
\centering 
 \caption{Critical dimensions for MBL with power laws} 
\begin{tabular}{c |c |c| c}
 \hline\hline &Unmixed & Anisotropic & Isotropic \\
&$\alpha = \beta$ \cite{Burin06} & $\beta < \alpha$ & $\beta < \alpha$ \\
\hline 
Hopping & $d < \alpha$ & $d < \alpha$    & $d < \alpha$  \\
Small Pairs  & $d < \beta$ & $d < \beta$ & $d < \beta+2$  \\
Extended Pairs & $d < \beta/2$ & $d < \frac{\alpha\beta}{\alpha+\beta}$  &$d < \frac{\alpha (\beta+2)}{\alpha +\beta +4}$ \\ 
Iterated Pairs & $d < \beta/2$  & $d < \beta/2$   & $d < (\beta+2)/2$ \\ \hline 

 \end{tabular} \label{table:critdim} 
\end{table}

As shown in Fig.~1a, the two strongly-hybridized central levels of a resonant pair define a new pseudo-spin degree of freedom (blue arrows) with local splittings $\delta \sim t/R_1^\alpha$.   
Pseudospins can  exchange energy through the interaction $V$ since the operators $S^z$ have spin-flip matrix elements between the two pseudospin states \cite{Burin06}.
Two pseudo-spins separated by $R_2$ resonate if $\delta_1,\delta_2>V(R_2) \gtrsim |\delta_1 - \delta_2|$ \cite{SuppInfo}. The number of such resonances available in a shell from distance $R_2$ to $2R_2$ around a fixed pseudospin (Fig.~1b) is 
\begin{align}
	\label{eq:pairpairnumber}
	N_2(R_1,R_2) \sim (n_1(R_1) R_2^d) \cdot \frac{V/{R_2^\beta}}{t/R_1^\alpha},
\end{align}
where $n_1 = \rho N_1$ is the density of pseudo-spins.
As before, if $N_2$ diverges as $R_2 \rightarrow \infty$, large scale pseudo-spin resonances induce delocalization \cite{Burin06}. 
There are two  limits.  The simplest case occurs when one holds the pair size $R_1$ fixed as $R_2$ diverges; this ``small pairs'' condition yields a localization criterion $d < \beta$. The second case requires optimizing $R_1$ as $R_2$ grows  in order to saturate the probability of pseudo-spin resonance. More precisely, one should replace $\frac{V/{R_2^\beta}}{t/R_1^\alpha} \rightarrow \text{min}[1,\frac{V/{R_2^\beta}}{t/R_1^\alpha}]$ in Eq.~\eqref{eq:pairpairnumber}.
The optimum arises for $R_1 \sim R_2^{\beta/\alpha}$,   yielding a more stringent ``extended pairs''  condition, $d < \frac{\alpha \beta}{\alpha +\beta}$.

\begin{figure}
\includegraphics[width=3.4in]{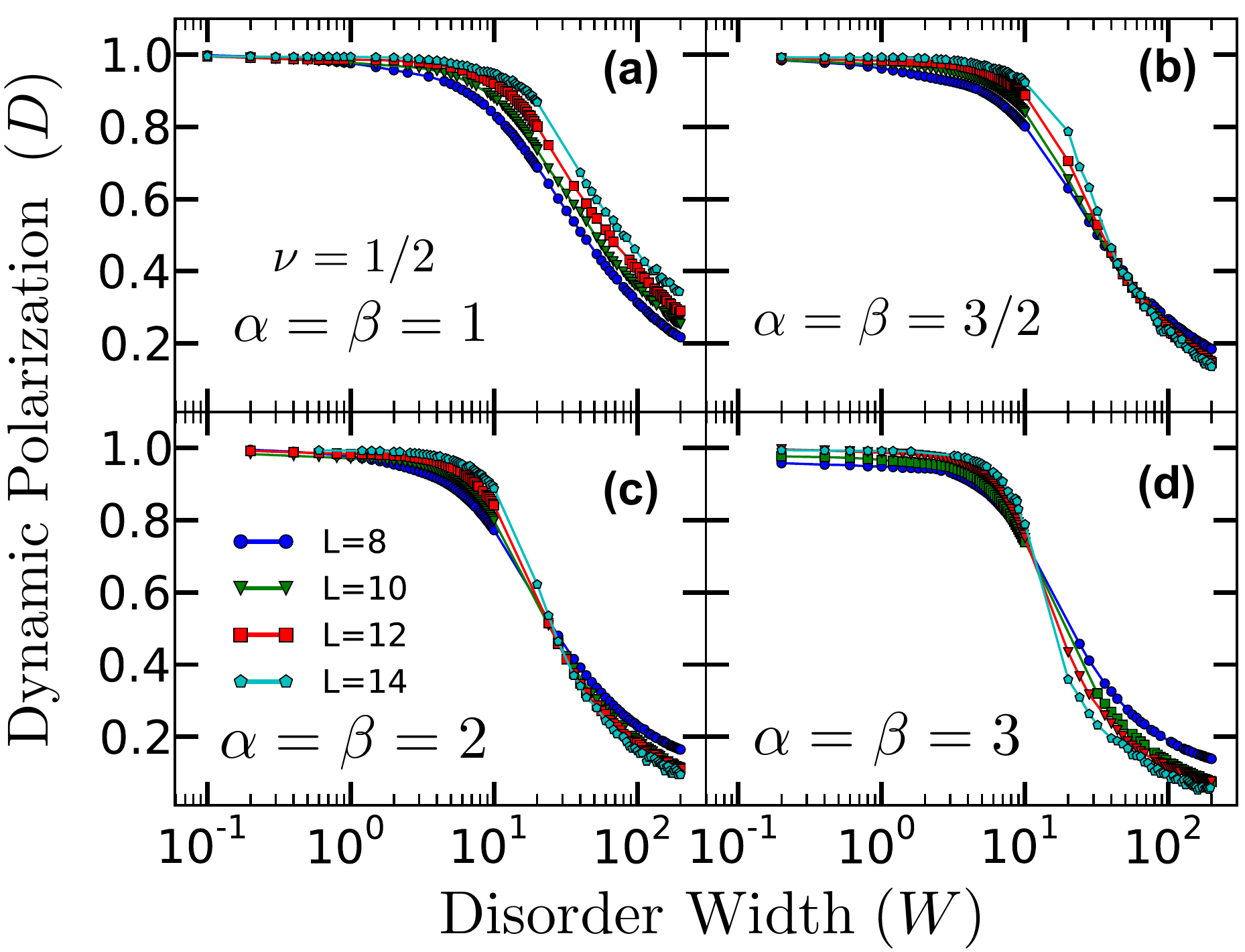}
\caption{%
\label{fig:fss_fracpolarization}
Finite size scaling of the long-time dynamic polarization for Eq.~\eqref{eq:hamgeneral} in $d=1$ (with units $t=1, V=2$) with a) $\alpha = \beta = 1$, b) $\alpha = \beta = 3/2$, c) $\alpha = \beta = 2$, and d) $\alpha = \beta = 3$.  The lack of flow reversal in (a) suggests delocalization at all disorders. The sharpening of the crossover as a function of increasing system size in (d) suggests the existence of a phase transition at approximately $W_c \approx 10$ into an MBL phase. The flow at intermediate power-laws (b) is inconclusive.   }
\end{figure}

It is clear that one can continue iterating the construction of pair resonances. However, the resulting criteria for MBL saturate after the third level \cite{SuppInfo, Burin07},
\begin{align}
	\label{eq:pairpairpairnumber}
	N_3(R_1,R_2,R_3) \sim ( n_2(R_1,R_2) R_3^d) \cdot \frac{V/{R_3^\beta}}{V/R_2^\beta} 
\end{align}
where $n_2 =  n_1 N_2$ is the density of pseudo-pseudo-spins. 
There are three  limits as $R_3$ diverges.  Holding $R_1,R_2$ fixed reproduces the small pairs criterion. Holding $R_1$ fixed but optimizing $R_2 \sim R_3$ (to saturate the probability of resonance in Eq.~\eqref{eq:pairpairpairnumber}) yields a new, ``iterated pairs'' criterion $d < \beta/2$. Finally, optimizing both $R_1 \sim R_2^{\beta/\alpha}$ and $R_2 \sim R_3$ reproduces the extended pairs criterion. 

The above results hold for generic anisotropic distributions of $t_{ij}, V_{ij}$ (first two columns of Table~I). In cases where the hoppings and interactions are isotropic, the effective matrix elements that arise in the four-spin construction cancel at leading order. This can be interpreted within a multipole expansion (for $R_1 < R_2$) which amounts to replacing $V/R_2^\beta \rightarrow VR_1^2/R_2^{\beta+2}$ for $N_2$ and analogously for subsequent iterations (last column of Table~I). 

A few comments are in order.  (1) In the anisotropic and unmixed ($\alpha =\beta$) cases, the iterated pairs criterion $d <\beta/2$ is always most stringent, a result first derived in \cite{Burin06}. (2) In the isotropic case, for $\alpha < \beta +4$, the extended pairs criterion is most stringent, while for $\alpha > \beta +4$ the iterated pairs criterion dominates. (3) The case of an Anderson insulator with Coulomb interactions corresponds to the $\alpha \rightarrow \infty$ limit of the isotropic case, giving an upper critical dimension of $d_c = 1.5$. (4) The case of interacting dipoles with $\alpha = \beta = 3$ also gives $d_c =1.5$. Interestingly, the orientation dependence of the dipolar interaction is sufficiently isotopic to enable a multipole expansion. Thus, in experiments that can realize $\alpha = 6$, $\beta = 3$ (as will be later discussed), $d_c \approx 2.3$ \cite{SuppInfo}.

Ultimately, all of the resonance arguments described above rely upon the analysis of  finite subsets of spins. While providing useful insights, such arguments must be viewed as heuristic. 
To supplement, we have performed extensive exact diagonalization studies of Eq.~\eqref{eq:hamgeneral} in $d=1$ for $\alpha = \beta = 1,3/2,2,3$.
We consider periodic systems up to size $L=14$ at filling fraction $\nu = 1/2$. The random fields are drawn from a uniform distribution of width $W$, the interaction $V_{ij} = V = 2$ and hopping $t_{ij} = t = 1$. 
The presence of a many-body localized phase may be detected by the finite size flow of the dynamic polarization $D$, a measure of spin transport across the 1D system at infinite temperature \cite{Pal10}. 
We perturb each eigenstate  with a small (long-wavelength) inhomogeneous spin modulation of the form $\hat{F} = \sum_j S^z_j e^{i 2\pi j/L}$ and measure the relaxation of this inhomogeneous polarization at infinite time.  
For each disorder realization $\eta$ and eigenstate $k$, the dynamic polarization is given by
\begin{equation}
D^k_{\eta} = 1 - \frac{ \langle k | \hat{F}^{\dagger} | k \rangle \langle k| \hat{F} | k \rangle}{ \langle k | \hat{F}^{\dagger} \hat{F} | k \rangle}.
\end{equation}
We then define $D$ as the infinite temperature disorder average of $D^k_\eta$. 
As $L \rightarrow \infty$, in the ergodic phase, one expects $D \rightarrow 1$ since the initial inhomogeneity relaxes away; in the MBL phase, one expects $D \rightarrow 0$ since there is no transport. 

The results are shown in Fig.~2. 
For all exponents, we find that the finite-size flow of $D$ is consistent with delocalization at weak disorder. 
At strong disorder, for $\alpha = 2, 3$ there are signs of flow reversal, consistent with a transition into an MBL phase, while for $\alpha=1$ the flow remains toward delocalization for all disorder strengths.
Owing to the small sizes accessible to exact diagonalization, flow reversal does not prove the existence of a transition; however, for $\alpha=3$ the combination of relatively clear flow and the previous theoretical argument suggests the existence of an MBL phase.
The strong disorder flow at intermediate exponents $\alpha = 3/2$ is inconclusive. 
Accordingly, for $d=1$, we numerically bound the critical power-law with $1< \alpha_c < 3$, noting that the extended pairs criterion gives $\alpha_c = 2$.  The difficulty of investigating an MBL transition in small size numerics emphasizes the importance of controlled experiments.


 \begin{figure}
\includegraphics[width=3.4in]{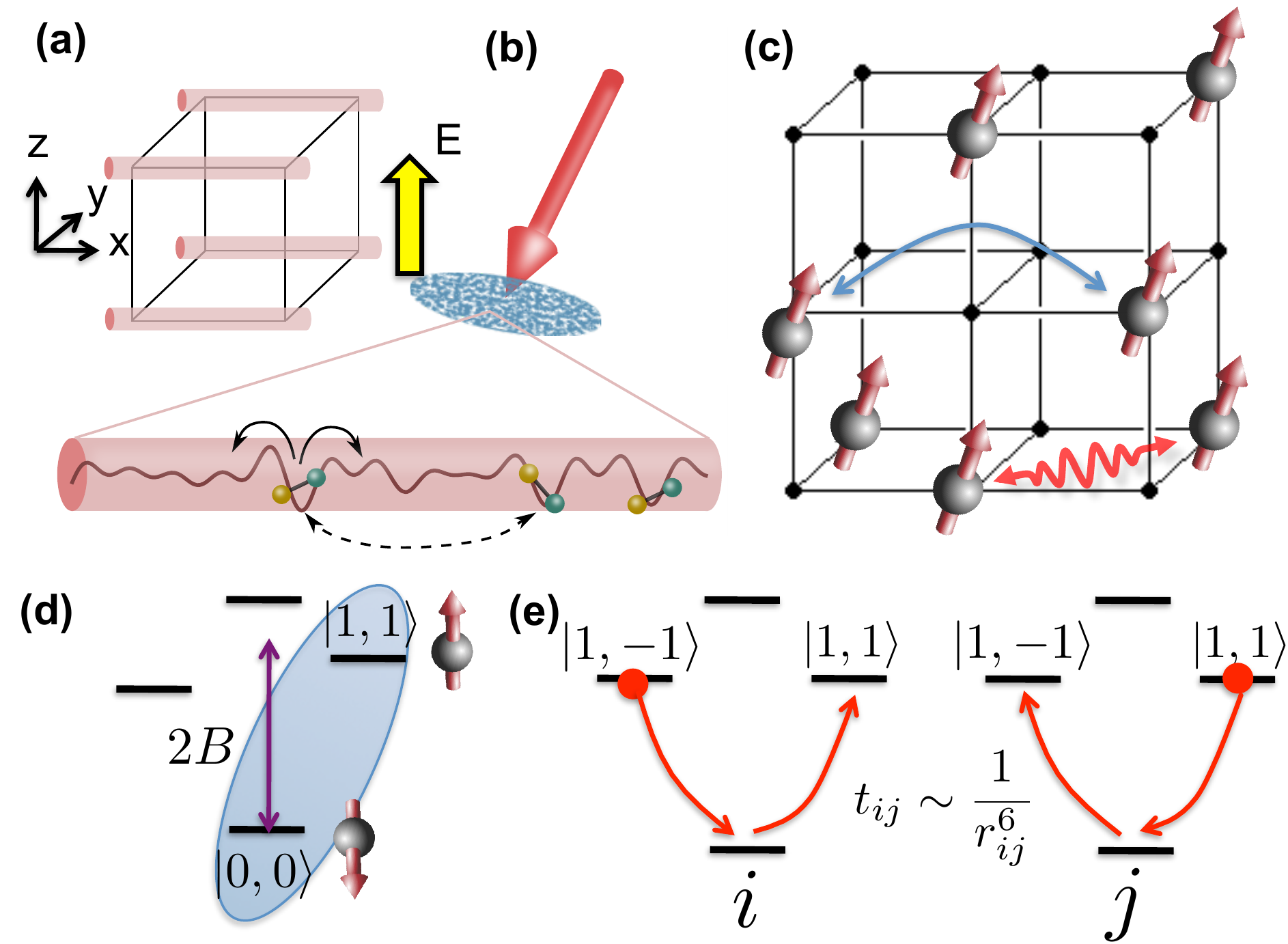}
\label{fig:tubes_pd}
\caption{%
a) Schematic of the one-dimensional tube geometry with strong confinement in the $\hat{y}$ and $\hat{z}$ directions and hopping in the $\hat{x}$ direction. 
b) Dipolar molecules in each 1D tube are subject to an optical speckle pattern which generates an effective random on-site chemical potential for the hopping molecules.
c) Schematic of the dipolar `spin' hopping model. Molecules pinned with dilution in deep optical lattice may exchange rotational excitations.
d) Effective rotational level structure of a polar molecule, with $\ket{\uparrow}=\ket{1,1}, \ket{\downarrow}=\ket{0,0}$ shown for the $\alpha=\beta=3$ rotor model.
e) Level structure of two polar molecules for the $\alpha=6$ rotor model, wherein hopping is mediated by a second order dipolar process.
}
\end{figure}

\paragraph{Experimental realizations ---}
We next analyze two classes of experimentally accessible systems in which MBL phases may be realized. 
First, we consider an array of polar molecules confined to a one-dimensional tube geometry (via an optical lattice) as depicted in Fig.~\ref{fig:tubes_pd}3a,b \cite{Miranda11}. 
The optical lattice is strongly confining along the $\hat{y}$ and $\hat{z}$ axes, but molecules can tunnel with nearest-neighbor hopping strength $t$ along the tube in the $\hat{x}$ direction ($\alpha \to \infty$).
The molecules are prepared in their rovibrational ground state and subject to a static electric field $E$ perpendicular to the tube direction. 
The applied electric field weakly aligns the molecules along the field direction, inducing a finite dipole moment $d$ and a long-range electric dipole-dipole interaction between the molecules $V \sim d^2/R^3$ ($\beta = 3$).
By ensuring that the dipolar interaction strength is much weaker than the rotational splitting $B$ (Fig.~3d), all molecules remain in the rovibrational ground state. 
Finally, an optical speckle field may be superimposed on top of the underlying  lattice introducing on-site potential disorder with strength $W$ controlled  by the  laser intensity (Fig.~3b) \cite{White09}.

The magnitude of the electric field  tunes the strength of the dipolar interaction $V \sim d^2$. In the limit $E \rightarrow 0$, the interaction strength $V\rightarrow0$, and the resulting nearest-neighbor Hamiltonian can be  fermionized. This non-interacting model is completely Anderson localized in the presence of any disorder. 
With the addition of local interactions, the existence of an MBL phase has been established both theoretically and numerically \cite{Basko06,Gornyi05,Oganesyan07,Pal10,Znidaric08}. 
According to the criterion in Table~I, the MBL phase ought to also survive the introduction of long-range dipolar interactions. To confirm this expectation and further establish an experimentally relevant phase diagram, we perform exact diagonalization  for molecular filling fractions $\nu = 1/2, 1/3, 1/4$ up to system sizes of $L=16, 18, 20$ respectively (Fig.~4a). 
As depicted in Fig.~4b, we obtain the MBL phase diagram as a function of interaction strength, filling fraction, and speckle intensity \cite{White09}.


We next consider disordered arrays of interacting molecules with fixed center of mass position and focus on the dynamics of rotational excitations (Fig.~3c).
In the deep lattice limit, the orbital motion of the molecules is pinned and the residual rotational degree of freedom is governed by an effective  Hamiltonian, $H_m= B J^2  - d^z E $ \cite{Brown03}. A combination of electric and magnetic  fields allows us to isolate an effective two-level system: $\left |\downarrow \right \rangle= |J=0,m_j=0 \rangle$ and $\left | \uparrow \right \rangle =|J=1,m_j=1 \rangle$ (Fig.~3d) \cite{Yan13}. 
The rotors interact via electric dipole-dipole interactions with Hamiltonian, $H_{dd} = \frac{1}{2}\sum_{i \neq j}  \frac{ {\bf{d_i}} (1-3 \hat{r}_{ij}  \hat{r}_{ij} ) {\bf  d_j}}{r_{ij}^3}$, where ${\bf d}$ is the dipole moment operator. Projecting $H_{dd}$ onto the two level subspace $\{ \ket{\downarrow},\ket{\uparrow} \}$ and keeping only secular terms yields the Hamiltonian of Eq.~\eqref{eq:hamgeneral} with effective on-site fields given by $\epsilon_i = \sum_{j\neq i} \frac{d_s d_a}{r_{ij}^3}$, $\alpha=\beta=3$,  and $d_{s,a} =\frac{ \langle 1| d^z | 1\rangle \pm\langle 0| d^z | 0\rangle}{2}$. Assuming Poissonian (uncorrelated) dilution, the fields $\epsilon_i$ become random variables with standard deviation $W \sim \frac{d_s d_a}{a_0^3} \sqrt{\nu(1-\nu)}$, where $a_0$ is the lattice spacing \cite{SuppInfo}. We expect the weak correlations of the random fields to leave the previous numerical phase diagrams in $d=1$  qualitatively unchanged (Fig.~2d). 

This dipolar spin model becomes particularly intriguing as one varies the dimensionality of the system since the ``extended pairs'' criterion predicts $d_c = 3/2$ for $\alpha = \beta = 3$. Compared to the simple Anderson criterion, which predicts $d_c = 3$, this already allows one to investigate the validity of the resonant pair  counting arguments for optical lattice pancakes where $d=2$.  

An additional feature of such systems is the ability to tune the spin-flip power-law. 
The large rotational constant $B$ enables restriction to the Hilbert space spanned by $\ket{\downarrow} = \ket{J=1, m_j=-1}$ and $\ket{\uparrow} = \ket{J=1, m_j=1}$. 
In this case, the dipolar flip-flop process is effectively eliminated at first order; the system instead hops two units of $J^z$ via a second order process of the form (Fig.~3e),
$
H'= \sum_{} \frac{t_{ij}^2}{r_{ij}^6}  \left[(d_+^{i})^2 (d_-^{j})^2 +  (d_-^{i})^2 (d_+^{j})^2\right]
$,
while the interaction remains formally unchanged. 
With the effective hopping power-law increased to $\alpha = 6$ and the interaction remaining as $\beta = 3$, one finds that (in $d=2$) all criteria for the consistency of localization are now satisfied, including both the extended pairs criterion which predicts $d_c \approx 2.3$ and the iterated pairs criterion with $d_c=2.5$.

\begin{figure}
\includegraphics[width=3.4in]{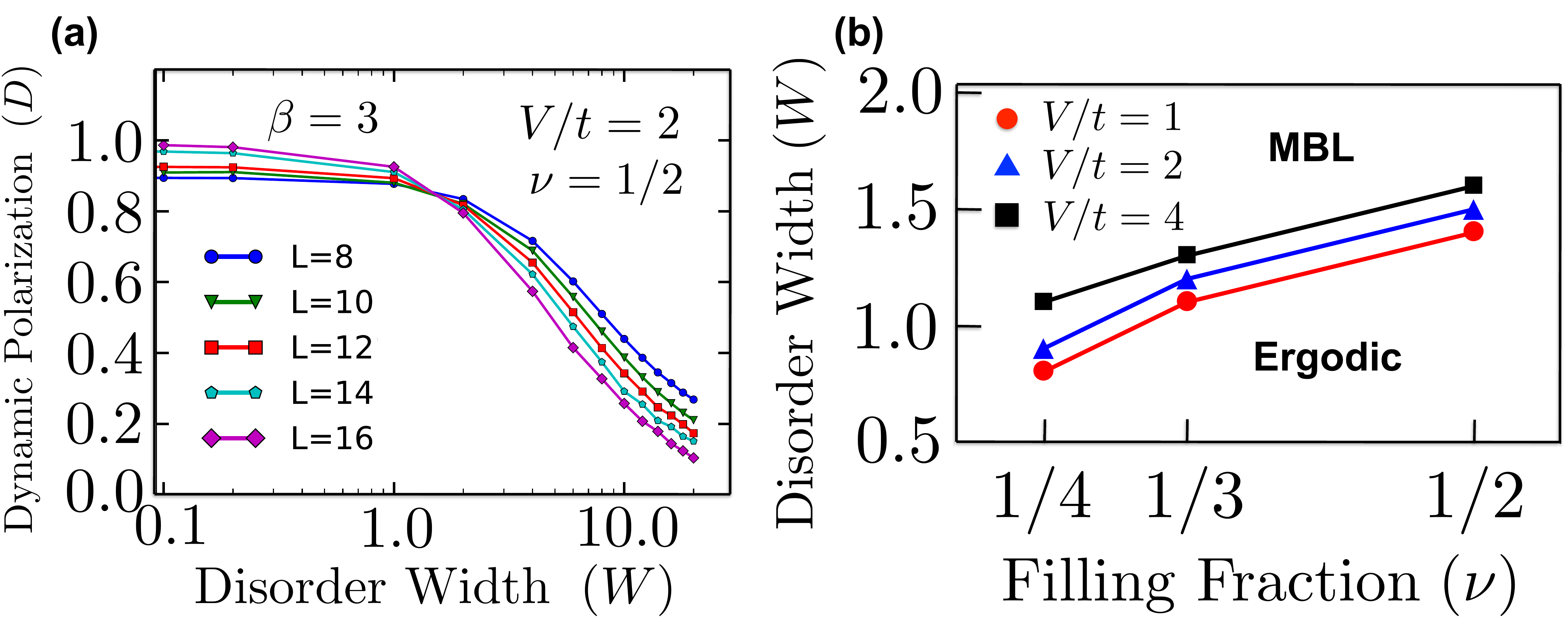}
\label{fig:fss_fracpolarization}
\caption{%
Exact diagonalization study of Eqn.~(1) with nearest neighbor hopping ($\alpha \rightarrow \infty$) and dipolar interactions ($\beta = 3$). Random fields are drawn from a uniform distribution of width $W$. 
a) Finite size scaling of the long-time dynamic polarization. The finite size flow suggests a delocalization phase transition  at approximately $W_c \approx 1.4t$. b)  MBL phase boundaries determined by finite-size flow for $V/t=1,2,4$. Error bars as determined by the width of the intersection region are smaller than markers. }
\end{figure}

Finally, solid-state implementations can be considered using spin defects in semiconductors. For example, Nitrogen-Vacancy (NV) defects in diamond \cite{Neumann10b, Childress06, Dutt07, Neumann08} are spin-1 magnetic impurities described by the Hamiltonian, $H_{NV} = D_0 S_z^2 + \mu_e B S_z$,
where $D_0$  is a large crystal field splitting. In the presence of an applied magnetic field, one can restrict the NV dynamics to a two-level subspace and  recover  
 the Hamiltonian of Eq.~\eqref{eq:hamgeneral}.

\emph{Experimental feasibility ---}
There are several probes available for detecting many-body localization in quantum optical systems: 1) observing arrested decay of a long-wavelength spin/number modulation, 2) generalized single-site spin-echo protocols that exhibit anomalously slow dephasing \cite{Bardarson12,Vosk12,Serbyn13,Huse13,Serbyn13b}, and 3) direct measurements of real-space correlation functions. The simplest approach is to directly observe a lack of diffusion. 
In a typical ergodic  system, an initial long-wavelength inhomogeneous number/spin polarization decays as $\sim e^{-Dk^2 t}$, where $D$ is the diffusion constant. 
For a many-body localized phase, $D = 0$. 
In any experiment,  coupling to an external bath is unavoidable and produces characteristic decoherence timescales; $T_1$-type depolarization provides a uniform $k$-independent contribution to the overall decay. 
In the presence of weak Markovian $T_2$ dephasing,  \emph{extrinsic} energy fluctuations induce diffusion, with $D_{T_2} \sim a_0^2 / T_2$ (neglecting back-action onto the bath).  
%
Since $T_2 \le T_1$, the figure of merit in such experiments is a separation of scale between $D_{T_2}$ and the expected ergodic diffusion, $D_e \sim a_0^2 / T_{a_0}$, where $T_{a_0}$ represents the lattice scale hopping time.
Alternatively, one can also measure the decay of an initially polarized region; for a Gaussian spot of initial size $\ell$ (larger than any correlation length), the modulation at the origin decays as $\sim (\ell^2 + Dt)^{-d/2}  e^{-t/T_1}$. Here, one hopes to extract the sub-exponential diffusive behavior, which can in principle be achieved by varying the spot size. 

%

In the molecular case, the most direct experimental realization of our proposals would be in diatomic alkali systems  \cite{ni08, Chotia12,Deiglmayr08,Aikawa10,Kerman04,Takekoshi12}.   Both the orbital and rotational cases can   be carried out  with currently available technology; indeed the loading of ${}^{40}$K${}^{87}$Rb molecules into 1D \cite{Miranda11} and 3D \cite{Chotia12} lattices, as well as  dipolar spin-exchange \cite{Yan13}, have  already been demonstrated. For a typical polar molecule with saturated dipole moment $ \sim 3$ Debye, the interaction strength at $532$nm (optical lattice spacing) corresponds to approximately $100$kHz, yielding $T_{a_0} \approx 10\mu$s. 
Meanwhile, dephasing times of up to $T_2\sim 100$ms \cite{Yan13} and ground-state lifetimes of up to $T_1\sim 25$s have been observed  \cite{Chotia12}.

%
%

In the case of NVs, recent advances in implantation and annealing have enabled dense  defect ensembles with average spacing $\sim 2-3$nm \cite{Maurer13}. The magnetic dipolar interaction at such distances is given by $T_{a_0} \sim 1\mu$s, significantly smaller than the typical room-temperature coherence times $T_1,T_2 \sim 10$ms of isolated NVs (working at cryogenic temperatures can lead to further improvements  \cite{Nir13}). To observe many-body localization in such a system will require the ability to reduce the effective dimensionality; this can be achieved by fabricating quasi-1D diamond nano-pillars \cite{Babinec10} or by controlled implantation in 2D layers \cite{Spinicelli11,Toyli10}. 

In summary, by constructing hierarchical spin resonances we have analyzed upper critical dimensions for many-body localization in the presence of power-laws (Table~I). Our arguments suggest that MBL is accessible to AMO-type experiments involving dipolar spins in two dimensions or hopping polar molecules in three or fewer dimensions. Our work opens a number of intriguing directions: (1) generalizations to other dipolar platforms such as Rydberg atoms, trapped ions and other spin qubits, (2)  working near the upper critical dimensions to probe the nature of the MBL transition.

It is a pleasure to gratefully acknowledge the insights of and discussions with A. Kamenev, D. Huse, A. Pal, R. Nandkishore, A. M. Rey, K. Hazzard, J. Ye, T. Pfau, A. Chandran, D. Abanin and A. Mirlin. This work was supported, in part, by the NSF, DOE (FG02-97ER25308), HQOC, Harvard-MIT CUA, the Lawrence Golub fellowship, the DARPA OLE program, AFOSR MURI, as well as the Austrian Science Fund (FWF) Project No. J 3361-N20.


\begin{thebibliography}{99}
\bibitem{footnotetij} $t_{ij}, V_{ij}$ do not scale with distance
\end{thebibliography}
\end{document}


\title{Supplemental Material for Many-body Localization with Dipoles}

\maketitle

\section{The Pseudo-spin Hamiltonian}

\begin{figure}[b]
\includegraphics[width=3.9in]{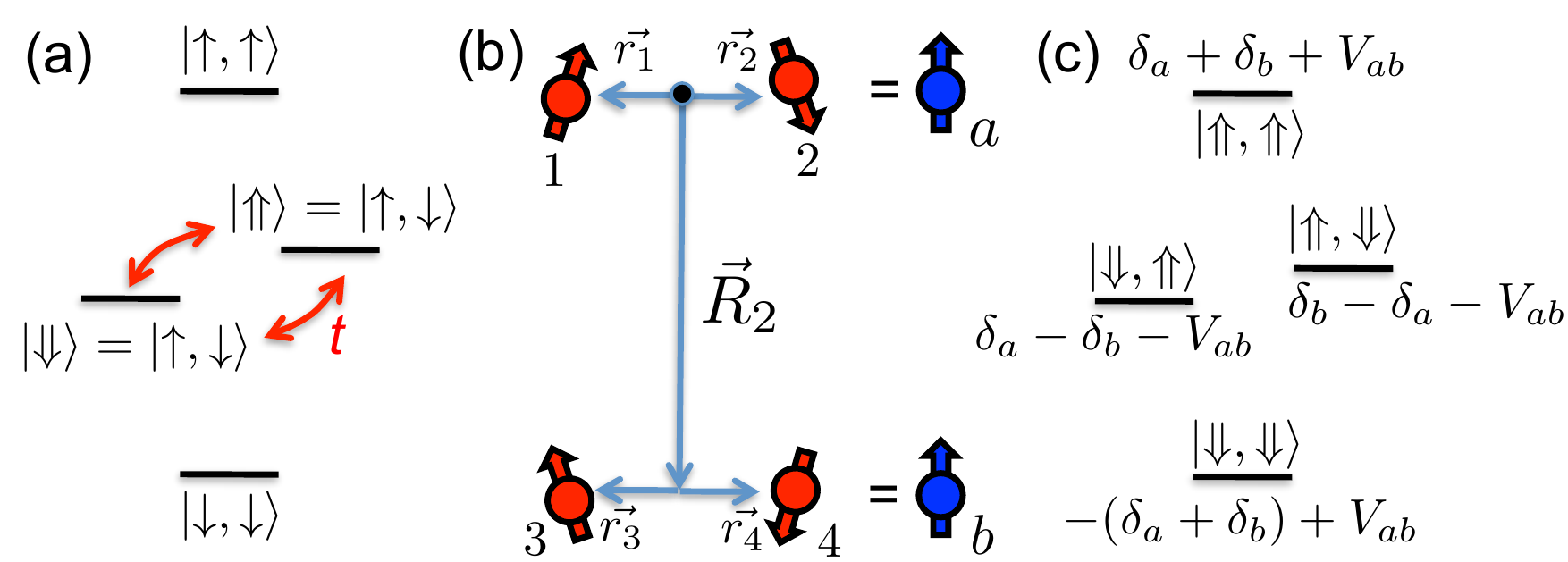}
\caption{Schematic construction of pseudo-spin resonances. (a) Energy levels of an original pair of spins separated by a distance $R_1$. The spin flip-flop couples the central two energy levels which defines a new pseudo-spin degree of freedom. (b) Spatial structure of an interacting pseudo-spin pair. The original spins $S_1$ and $S_2$ form the new pseudo-spin $a$ and spins $S_3$ and $S_4$ form the new pseudo-spin $b$. (c) Bare energy levels (neglecting $t_a, t_b$) of the pseudo-spin Hamiltonian $H_{ab}$.  }
\label{fig:YSRscheme}
\end{figure}

 We begin by isolating the Hamiltonian of a pair of spins, $S_1, S_2$, at some separation $R$,
\begin{align}
	H_{12} = \epsilon_1 S^z_1 + \epsilon_2 S^z_2 + t_{12} (S^+_1 S^-_2 + S^-_1 S^+_2) + V_{12} S^z_1 S^z_2
\end{align}
where $\epsilon_i$ are local random fields of bandwidth $W$, $t_{12}, V_{12}$ are the flip-flop and interaction piece of the Hamiltonian and we have temporarily absorbed the $R$-dependence into the couplings. 
For $R$ large enough, we may assume that $ t_{12}, V_{12} \ll |\epsilon_i| \sim W$.
In this case, the perturbation $t_{12}$ leads to resonance between the $\ket{\Uparrow} = \ket{\uparrow\downarrow}$ and $\ket{\Downarrow} = \ket{\downarrow\uparrow}$ states in the $S^z = 0$ manifold, precisely when the detuning $\delta_a = \epsilon_1 - \epsilon_2$ satisfies $|\delta_a| \lesssim t_a = t_{12}$. 
Somewhat more formally, we define a set of pseudospin Pauli operators $\tau_a^\alpha$ with respect to the $\ket{\Uparrow}, \ket{\Downarrow}$ basis to get an effective pseudospin Hamiltonian restricted to the $S^z = 0$ subspace:
\begin{align}
	H_a = \delta_a \tau_a^z + t_a \tau_a^x
\end{align}
Resonance corresponds to the condition that the eigenstates of this Hamiltonian point predominantly in the $\tau_a^x$ direction.
We note that $V_{12}$ does not enter the pseudospin Hamiltonian as the Ising term $S^z_1 S^z_2$ is constant on the $S^z = 0$ subspace.

 We now consider the interaction of a pair of pseudospins $a=12$, $b=34$ each of size $R_1$ and at separation $R_2$ (Fig.~S1). So long as $t(R_2) \ll |\epsilon_i| \sim W$, the hopping terms between $a$ and $b$ spins will be unable to resonantly move out of the $S^z_a = 0$, $S^z_b = 0$ subspaces, so we may again restrict attention to the joint pseudospin space. The effective Hamiltonian in this space is
\begin{align}
	H_{ab} = \delta_a \tau_a^z + t_a \tau_a^x + \delta_b \tau_b^z + t_b \tau_b^x + 
	V_{ab} \tau_a^z \tau_b^z
\end{align}
where $V_{ab} = V_{13} - V_{14} - V_{23} + V_{24}$. This coupling formula follows most easily by noting that, within the pseudospin subspace, $\tau_a^z = S_1^z = - S_2^z$.

 Within the effective Hamiltonian $H_{ab}$, pseudospin resonance corresponds to the condition that some set of eigenstates of $H_{ab}$ are entangled of order unity (as a function of the separations $R_1,R_2$). 
Clearly this requires the interaction $V_{ab} \ne 0$, but it further requires $t_{a/b} \ne 0$ or else the eigenstates will be product states in the $\tau^z$ basis. 
Assuming the original spins are resonant ($\delta_{a/b} \lesssim t_{a/b}$), we find pseudospin resonance for $\sqrt{t_a^2+\delta_a^2}, \sqrt{t_b^2+\delta_b^2} \gtrsim V_{ab} \gtrsim |t_a - t_b|$. 
This corresponds to the resonance condition quoted (above Eqn.~3) in the main text.

 Finally, we turn to the next level of the hierarchical construction of resonances, \emph{i.e.} pseudo-pseudo-spins [Eqn.~(5) main text]. We define a new set of pseudo-pseudo-spin Pauli operators $\mu^{\alpha}$ and take the $\mu^z$ eigenstates to label the  two resonant central eigenstates  of $H_{ab}$. Unlike the pseudospins $\tau$, there is no natural quantization axis for  $\mu$. In particular, that the two central eigenstates are resonant ensures that the  pseudo-pseudo-spins have $\mathcal{O}(1)$ transition matrix elements with respect to the underlying $\tau^x$ and $\tau^z$ operators. Thus, a pair of pseudo-pseudo-spins has a Hamiltonian of the form,
\begin{align}
	H &= H_{ab} + H_{cd} + H_{int} \nonumber \\  
	&= \Delta_{ab} \mu^z_{ab} + \Delta_{cd} \mu^z_{cd} + \sum_{\alpha,\beta \in \{x,z\}}V_{\alpha \beta} \mu^{\alpha}_{ab} \mu^{\beta}_{cd},
\end{align}
where $H_{int} = V_{ac}\tau^z_a \tau^z_c + V_{ad}\tau^z_a \tau^z_d+V_{bc}\tau^z_b \tau^z_c+V_{bd}\tau^z_b \tau^z_d$ and in the second line, we have  restricted to the resonant  $\mu$ subspace.

\section{Multipole expansion} 
\label{sec:multipole_expansion}

 The pseudospin resonance between two well-separated pairs of spins corresponds to correlated interaction-induced local charge rearrangements within each pair. If $V(r) \sim 1/r$ is pure Coulomb, this observation immediately suggests that the effective interactions between pairs should decay according to the next leading term in a multipole expansion --- that is, as dipoles $1/r^3$. More generally, for homogenous, isotropic interactions $V(r) \sim 1/r^\beta$, we have
\begin{align}
	V_{ab} &= V_{13} - V_{14} - V_{23} + V_{24} \nonumber \\ 
	& =  \left (\frac{1}{R_{13}^\beta }  - \frac{1}{R_{14}^\beta }    \right )   + \left (\frac{1}{R_{24}^\beta }  - \frac{1}{R_{23}^\beta }    \right )  \nonumber \\
	&= \left (\frac{1}{| \vec{R}_2 + \vec{r_4} - \vec{r_1}|^{\beta}}  -\frac{1}{| \vec{R}_2 + \vec{r_3} - \vec{r_1}|^{\beta}}   \right )   + \left (\frac{1}{| \vec{R}_2 + \vec{r_3} - \vec{r_2}|^{\beta}}  -\frac{1}{| \vec{R}_2 + \vec{r_4} - \vec{r_2}|^{\beta}}    \right )  \nonumber \\
	& \approx  \frac{1}{R_2^\beta} \frac{\beta}{2}\left (        \frac{2 r_3 \cdot r_1 + 2 r_4 \cdot r_2 -2 r_4 \cdot r_1 -2 r_3 \cdot r_2 }{R_2^2} \right) \sim \frac{R_1^2}{R_2^{\beta+2}}.
	\label{eq:multipole_form}
\end{align}
where in the last step we have assumed that $R_1 < R_2/2$ in order to perform the multipole expansion. The factor of $R_1^2$ in the numerator corresponds to the scale of the multipole moment.

 In general, the cancellation that produces the leading multipole decay fails unless the interactions are homogeneous and isotropic. An important special case is provided by uniformly aligned dipoles for which the interaction depends on the angle between the dipole axis and the displacements $R_{ij}$. 
The sum $V_{ab} = V_{13} - V_{14} - V_{23} + V_{24}$ can be reinterpreted as the interaction energy of the four dipoles where the dipole at site 2 and 4 has been reversed and thus the net dipole moment in $a$ or $b$ is zero and $V_{ab} \sim R_1^2/R_2^5$ becomes quadrupolar.

 When the multipole form of $V_{ab}$ Eq.~\eqref{eq:multipole_form} applies, the estimate of $N_2(R_1,R_2)$ presented in the main text Eq.~(3) must be corrected by the replacement $V/{R_2^\beta} \rightarrow V R_1^2 / {R_2^{\beta+2}}$. 
The resulting criteria for MBL (the ``istotropic'' case) are summarized in the last column of Table~I. For completeness, we provide detailed derivations of these formulae below.


\section{Pseudo-spin counting}

\subsection{Fixed pair size (``small pairs'')}

 We fix the size of a central pseudo-spin to be $R_1$ and count the number of resonant pseudo-spins  separated by a distance $R_2$ (using the multipole form of the interaction), 
\begin{align}
	\label{eq:n2_form}
N_2(R_1,R_2) \sim   (n_1(R_1) R_2^d) \cdot \frac{V R_1^2/R_2^{\beta+2}}{t/R_1^\alpha} \sim R_2^{d-(\beta+2)},
\end{align}
where $n_1 = \rho N_1$ is the density of pseudo-spins. The first factor ($n_1(R_1) R_2^d$) counts the total number of pseudo-spins in a volume shell between $R_2$ and $2R_2$ while  the second factor ($\frac{VR_1^2/R_2^{\beta+2}}{t/R_1^\alpha}$) represents the probability that a given pseudospin is resonant with the central pseudospin. With $R_1$ fixed, we can take $R_2 \rightarrow \infty$ to check whether the number of resonant pseudo-spins diverges. This  occurs when $d > \beta +2$.

\subsection{Growing pair size (``extended pairs'')}

 A more stringent constraint arises when one allows the size of the central pseudospin to grow as $R_2$ grows.
The optimum arises when $V(R_2) \sim t(R_1)$ such that the factor describing the probability of resonance  in Eq.~\eqref{eq:n2_form} is maximized (of order unity), $\frac{V(R_2)}{t(R_1)}\sim \frac{R_1^2/R_2^{\beta+2}}{1/R_1^\alpha} \sim 1/10$, yielding $R_2 \sim R_1^{\frac{\alpha +2}{\beta+2}}$ (again using the multipole form of the interaction) and 
\begin{align}
N_2(R_1,R_2)  \sim R_1^{d +2} (R_1^{\frac{\alpha +2}{\beta+2}})^{d-(\beta+2)} = R_1^{d-\alpha +d \frac{\alpha +2}{\beta+2}}.
\end{align}
The number of resonant pseudo-spin pairs diverges at large scales, and hence delocalization occurs, for $d > \frac{\alpha (\beta+2)}{ \alpha+\beta +4}$. 

\subsection{Iterating the construction of pseudo-spin pairs (``iterated pairs'')}

 It is possible to continue iterating the hierarchical construction of resonant pairs (e.g. to create an effective pseudo-pseudo-spin from 4 original spin degrees of freedom). However, the resulting criteria for MBL saturate after the third level. The counting at this level is,
\begin{align}
	\label{eq:pairpairpairnumber}
	N_3(R_1,R_2,R_3) \sim ( n_2(R_1,R_2) R_3^d) \cdot \frac{\tilde{V}/{R_3^\beta}}{\tilde{V}/R_2^\beta} =( n_2(R_1,R_2) R_3^d) \cdot \frac{\tilde{V}R_1^2/{R_3^{\beta+2}}}{\tilde{V}R_1^2/R_2^{\beta+2}} = R_1^{2d -\alpha +2} \cdot R_2^{d} \cdot R_3^{d-\beta-2} 
\end{align}
where $n_2 =  n_1 N_2$ is the density of pseudo-pseudo-spins. 
As usual, we count the number of resonant pseudo-pseudo-spins as $R_3 \rightarrow \infty$. Holding $R_1,R_2$ fixed reproduces the small pairs criterion from Eqn.~S5.
Holding $R_1$ fixed but optimizing $R_2 \sim R_3$ (to saturate the probability of resonance in Eqn.~S7) yields a new, ``iterated pairs'' criterion $d > (\beta+2)/2$. Finally, optimizing both $R_1\sim R_2^{(\beta+2)/(\alpha+2)}$ and $R_2\sim R_3$ reproduces the extended pairs criterion of Eqn.~S6.  Physically, the reason that all MBL criteria saturate after $N_3$ is that no new length scales emerge; this occurs because both the numerator and denominator of the term describing the  probability of resonance  originate from $V$ as all pseudo-pseudo-spins have transition dipole moments with respect to $\sigma^z$. The iterated pairs criterion supersedes Eq.~(S6) only when $\alpha > \beta+4$. We note that throughout section III we always assume  $R_1\le R_2\le R_3$.

\section{Nearby resonances} 
\label{sec:nearby_resonances}

 In the mixed power law regime, with $\alpha > \beta$, the distance $R_2$ between pairs of pseudospins is much larger than the scale of the both isotropic ($R_1 \sim R_2^{\beta+2/\alpha+2}$) and anisotropic ($R_1 \sim R_2^{\beta/\alpha}$) pseudospins.
 As the interactions V are generally much stronger, one might worry that pseudospins at distances $\tilde{R} < R_2$ might spoil the resonance condition for the pseudospin counting at  the larger distance $R_2$.

 When are there $\mathcal{O}(1)$ pseudospins of scale $R_1$ at distances $\tilde{R} < R_2$? First, we note that the number of resonances at a distance $\tilde{R}$ is given by,
\begin{align}
N_2(R_1,\tilde{R})  \sim \rho N_1(R_1) \tilde{R}^d \sim R_1^{d-\alpha} \tilde{R}^d.
\end{align}
Thus, there are $\mathcal{O}(1)$ resonances when $\tilde{R} \sim R_1^{\alpha/d-1}$. We should now compare this with $R_2$ as set by the condition $V(R_2) \sim t(R_1)$. There are three possible cases:
\begin{enumerate}
	\item  When $\tilde{R} > R_2$, there are no ``nearby'' resonances.
	\item When $\tilde{R} = R_2$, pairs at scale $R_1$ find other pairs at scale $\tilde{R} = R_2$ which are resonant with respect to $V$.
 	\item When $\tilde{R} < R_2$, there are many ``nearby'' resonances before distance $R_2$.
\end{enumerate}

 The critical case occurs when $\tilde{R} \sim R_2$ or when $R_1^{\alpha/d-1} \sim R_1^{\frac{\alpha}{\beta}}$,  yielding $d_c = \frac{\alpha \beta}{ \alpha+\beta }$. This result is especially nice since this condition matches with that obtained for the critical dimensions for MBL (middle column of Table~I in the maintext). Thus, for $d <  \frac{\alpha \beta}{ \alpha+\beta }$, MBL is consistent with resonance counting of the form considered above; moreover, there are no ``nearby'' resonances to modify the counting. 


\section{Size of a resonant pseudo-spin pair} 
\label{sec:nearby_resonances}

 In numerical and experimental studies of finite size systems,  pseudospin resonances only play a role when the system size becomes large enough to contain them. Thus, in this section, we estimate the typical scale $R_2$ at which any given extended pair finds $\mathcal{O}(1)$ resonant partners $N_2(R_2) \sim 1$. The typical size $R_2$ is measured in lattice units and thus has a microscopically detailed dependence on the microscopic couplings and disorder as we derive below. 

\subsection{Random field disorder}
 Let us first estimate $R_2$ in the case of a filled lattice where disorder arises from the underlying randomness of the on-site fields (e.g. the case of molecules with speckle). In the next subsection, we will consider the case where disorder arises from dilution (e.g. rotor or NV spin model).

 We consider $N_2(R_1,R_2)$ and are interested in the smallest $R_2$ which harbors resonances. For concreteness, let us work with $\alpha =\beta$ and hence take $R_1 \sim R_2$. A central pseudo-spin is resonates with another pseudo-spin at scale $R_2$ if $N_2(R_2) \sim 1$, so we estimate,
\begin{align}
	\label{eq:pairpairnumber}
	 1 \sim N_2(R_2) =  \rho^2 \cdot R_2^{2d-\beta } \cdot \frac{V}{W} \implies R_2^{2d-\beta} = \frac{1}{\rho^2} \frac{W}{V} = \frac{1}{\rho^2} \frac{1}{a_0^\beta}  \frac{W}{V/a_0^\beta} 
\end{align}
We now define  $D = \frac{W}{V/a_0^\beta}$ as the ratio between the disorder bandwidth and the nearest neighbor interaction strength; it is a dimensionless measure of the disorder strength and  for an un-diluted lattice $\rho \sim  1/a_0^d$. Thus, one obtains
\begin{align}
	\label{eq:pairpairnumber}
	 R_2 \approx a_0 D^{1/(2d-\beta)},
\end{align}
which for $d=\beta=3$ is $R_2 \approx a_0 \sqrt[3]{D}$ and for $d=2$ is $R_2 \approx a_0 D$. For $\alpha = \beta = 3$ (e.g. dipoles), the expected upper critical dimension is $d_c =1.5$ so as expected, the pair size diverges (in large disorder, such that $D>1$) as one approaches $d_c$ from above.

\subsection{Disorder arising from dilution} 
\label{sec:nearby_resonances}

 We now consider the case where disorder arises from Poissonian (random) dilution on the lattice.  For concreteness, let us consider the rotor example from the maintext ($\alpha = \beta = 3$). The effective on-site random fields generated by the dilution  arise from the $d^z_i d^z_j$  (Ising) term of the dipolar interaction.  Assuming that the rotors are oriented along the same quantization axis, one finds
\begin{align}
	\label{eq:hamgeneral}
	\sum_{i < j} \frac{d^z_i d^z_j}{r_{ij}^3} = \sum_{i < j} \frac{(d^s_i + d^a_i \sigma^z_i) (d^s_j + d^a_j \sigma^z_j)}{r_{ij}^3} = \sum_{i < j} \frac{(d^s)^2}{r_{ij}^3} +  \sum_{i < j} \frac{d^a d^s}{r_{ij}^3} (  \sigma^z_i + \sigma^z_j) + \sum_{i < j} \frac{(d^a)^2}{r_{ij}^3} \sigma^z_i  \sigma^z_j,
\end{align}
where we have re-expressed the effective permanent dipole moment (diagonal component of the dipole moment operator) in terms of symmetric and antisymmetric pieces, e.g.~$ \bra{ \uparrow} d^z \ket{\uparrow} =d^s + d^a$ and $ \bra{ \downarrow} d^z \ket{\downarrow} =d^a - d^s$.

 Let us now estimate the width of the disorder distribution arising from random fields $\epsilon_i = \sum_{i \neq j} \frac{d^a d^s}{r_{ij}^3}$.  The average field at site $i$ is $\overline{\epsilon_i} = \sum_{j \neq i}  \frac{d^a d^s}{r_{ij}^3} \overline{Q_j} = \nu   \frac{d^a d^s}{a_{0}^3} \sum_{\ell \in lat} \frac{1}{\ell^3}$,
where $\nu$ is the dilution probability and  $Q_j=1$ for a lattice site containing a spin and $0$ otherwise; the last sum runs over the integer lattice and is the analog of the usual Madelung constant in electrostatics. The variance of the on-site field is $\overline{\delta \epsilon_i^2} = \overline{\epsilon_i^2} -  \overline{\epsilon_i}^2$, where $ \overline{\epsilon_i^2}  = \overline{  \sum_{j \neq i} \sum_{k \neq i} \left ( \frac{d^a d^s}{r_{ij}^3} Q_j \right) \left( \frac{d^a d^s}{r_{ik}^3} Q_k \right) }$.
Noting that $\overline{Q_j Q_k} = \nu \delta_{jk} + \nu^2 (1-\delta_{jk})$ yields $
\overline{ \epsilon_i^2 } = \left ( \frac{d^a d^s}{a_{0}^3} \right)^2 \left [ (\nu-\nu^2) \sum_{\ell \in lat} \frac{1}{\ell^6} + \left(\nu \sum_{\ell \in lat} \frac{1}{\ell^3}\right)^2  \right  ]$.
Thus, one finds that 
\begin{align}
	\label{eq:hamgeneral}
	W= \sqrt{\langle \overline{ \epsilon_i^2 }} =\sqrt{ \overline{ \epsilon_i^2 } - \overline{ \epsilon_i}^2 }= \frac{d^a d^s}{a_{0}^3} \sqrt{   (\nu-\nu^2) \sum_{\ell \in lat} \frac{1}{\ell^6}   },
\end{align}
which in the strong dilution limit ($\nu \ll 1$) scales as $W \sim \nu \frac{d^a d^s}{a_{0}^3}$.

 With the effective disorder bandwidth in hand, we now estimate the size of resonant pseudo-spin pairs. For the case of dilution,  we note that $\rho \sim \nu/a_0^3$ and  $W= \frac{d^a d^s}{a_{0}^3} \sqrt{   (\nu-\nu^2) \sum_{\ell \in lat} \frac{1}{\ell^6}   }$.  Following the same analysis, one finds,
\begin{align}
	\label{eq:pairpairnumber}
	 1 \sim N_2(R_2) =  \rho^2 \cdot R_2^{2d-\beta} \cdot \frac{V}{W} \implies R_2^{2d-3} = \frac{1}{\rho^2} \frac{W}{V} = \frac{1}{\rho^2} \frac{1}{a_0^3}  \frac{ \frac{d^a d^s}{a_{0}^3} \sqrt{   (\nu-\nu^2) \sum_{\ell \in lat} \frac{1}{\ell^6}   } }{V/a_0^3}.
\end{align}
The interaction strength $V \sim (d^a)^2$ and hence,
\begin{align}
	\label{eq:pairpairnumber}
	 R_2 \approx a_0 \left ( \frac{d^s}{d^a} \frac{1}{\nu^2} \sqrt{   (\nu-\nu^2) \sum_{\ell \in lat} \frac{1}{\ell^6}   } \right )^{\frac{1}{2d-3}}.
\end{align}
Working in the strong dilution limit $\nu \ll 1$ and ignoring the order 1 correction from the Madelung constant, one finds that in $d=3$, $R_2 \approx \left (\frac{d_s}{d_a} \right )^{1/3} a_0 / \sqrt{\nu}$ and in $d=2$, $R_2 \approx \frac{d_s}{d_a}  a_0 / \nu^{3/2}$.